\documentclass[10pt,journal,a4paper,final,oneside,twocolumn]{IEEEtran}

\usepackage{times}
\usepackage{graphicx}
\usepackage{float}
\usepackage{cite}
\usepackage{color}
\usepackage{pifont}
\usepackage{amsmath,amssymb,amsfonts}
\usepackage{enumerate}
\usepackage{subfigure}
\usepackage{multirow}
\usepackage{diagbox}
\usepackage[top=0.6in, bottom=0.6in, left=0.5in, right=0.5in]{geometry}

\renewcommand{\baselinestretch}{1}
\begin{document}
\setlength{\abovedisplayskip}{1pt}
\setlength{\belowdisplayskip}{1pt}

\title{Polar Coded Integrated Data and Energy Networking: A Deep Neural Network Assisted End-to-End Design}
\author{Luping Xiang, {\em Member, IEEE}, Jingwen Cui, Jie Hu, {\em Senior Member, IEEE}, Kun Yang, {\em Fellow, IEEE}, Lajos Hanzo, {\em Life Fellow, IEEE}
\thanks{The authors would like to thank the financial support of Natural Science
Foundation of China (No. 62132004, 61971102), MOST Major Research
and Development Project (No. 2021YFB2900204), Sichuan Science and
Technology Program (No. 2022YFH0022, 22QYCX0168), UESTC Yangtze Delta Region
Research Institute-Quzhou (No. 2022D031), International Postdoctoral Exchange Fellowship Program (No. YJ20210244) and EU H2020 Project COSAFE (GA-824019).}
\thanks{L. Hanzo would like to acknowledge the financial support of the 
Engineering and Physical Sciences Research Council projects EP/W016605/1 
and EP/X01228X/1 as well as of the European Research Council's Advanced 
Fellow Grant QuantCom (Grant No. 789028)}
\thanks{Luping Xiang, Jingwen Cui, Jie Hu and Kun Yang are with the School of Information
and Communication Engineering, University of Electronic Science and
Technology of China, Chengdu 611731, China, email: luping.xiang@uestc.edu.cn, 202221011120@std.uestc.edu.cn, hujie@uestc.edu.cn. \textit{(Corresponding author: Jie Hu.)}}
\thanks{Kun Yang is also with the School of Computer Science and Electronic Engineering,
University of Essex, Essex CO4 3SQ, U.K., e-mail: kunyang@essex.ac.uk}
\thanks{Lajos Hanzo is with the School of Electronics and Computer Science, University of
Southampton, Southampton SO171BJ, U.K., e-mail: lh@ecs.soton.ac.uk}
}

\maketitle

\begin{abstract}
Wireless sensors are everywhere. To address their energy supply, we proposed an end-to-end design for polar-coded integrated data and energy networking (IDEN), {where the conventional signal processing modules, such as modulation/demodulation and channel decoding, are replaced by deep neural networks (DNNs)}. Moreover, the input-output relationship of an energy harvester (EH) is also modelled by a DNN. By jointly optimising both the transmitter and the receiver as an autoencoder (AE), we minimize the bit-error-rate (BER) and maximize the harvested energy of the IDEN system, while satisfying the transmit power budget constraint {determined by the normalization layer in the transmitter}. 
Our simulation results demonstrate that the DNN aided end-to-end design conceived outperforms its conventional model-based counterpart both in terms of the harvested energy and the BER.
\end{abstract}

\vspace{-0.5 cm}
\section{Introduction}
Wireless sensors are becoming pervasive in support of the Internet of Everything (IoE) \cite{hu2020energy}. However, their limited energy storage constrains their operational cycles. Fortunately, radio-frequency (RF) signals can be relied upon for controllable wireless energy transfer (WET) towards these miniature sensors. {Generally, the RF signals simultaneously convey energy as well as information, which forms the basis of integrated data and energy networking (IDEN). The WET aims to meet the associated recharging requirement, while the wireless information transfer (WIT) aims for meeting the communication requirement.}
But again, coordinating both WIT and WET within the same spectrum is challenging, although highly desirable for simultaneously satisfying both communication and recharging requirements \cite{perera2017simultaneous}.

The concept of end-to-end communication system was proposed in \cite{o2017introduction} for improving the attainable performance in complex scenarios, in the face of uncertainties where conventional mathematical methods were hard to apply. In such systems, the transmitter, the channel, and the receiver may be implemented in the form of deep neural networks (DNNs), which can be trained together as an autoencoder (AE). This approach does not rely on the classical functional modules for modulation/demodulation, hence it is also often termed as being model-free. This novel architecture achieves competitive bit error rate (BER) performance, when compared to traditional model-based communication system. This is because the DNN aided model-free transceiver is capable of jointly optimizing the entire process from the generation of data bits at the transmitter to their reception at the receiver, which constitutes a so-called ``end-to-end'' design. Moreover, this design allows the transceiver to cope with the imperfections of practical systems, such as their non-linearity for example. Since the  channel is unknown in practice, Aoudia and Hoydis \cite{aoudia2019model} presented a new learning algorithm, which alleviated this problem by training the transmitter and receiver differently. Explicitly, they trained the receiver with the aid of an approximation of the loss function gradient, while training the transmitter by relying on the true gradient. Most AEs were trained based on the symbol-level information \cite{aoudia2019model, jiang2022residual, varasteh2020learning}, but this philosophy is incompatible with practical bit-metric based decoding (BMD) at the receivers \cite{9713687}. Therefore, Cammerer \textit{et al.} \cite{cammerer2020trainable} conceived an AE based on bit-wise mutual information (BMI), which was eminently suitable for integration with practical receivers.

\begin{table}[t]
\centering
\caption{Contrasting Our Contributions To The State-Of-Art}
\begin{tabular}{l|c|c|c|c|c } 
 \hline
 Contributions & \textbf{this work} & \cite{o2017introduction, aoudia2019model, jiang2022residual} & \cite{cammerer2020trainable} & \cite{varasteh2020learning} & \cite{9109744zhu,8109997,teng2019low} \\ \hline\hline
 Autoencoder & \ding{52} & \ding{51} & \ding{51} & \ding{51} &  \\ 
 \hline
 {Based on BMI} & \ding{52} &  & \ding{51} &  &  \\ 
 \hline
 DNN-adied EH & \ding{52} &  &  & \ding{51} &   \\
 \hline
 Polar decoder & \ding{52} &  &  &  & \ding{51}  \\
 \hline
 Joint optimization & \ding{52} &  &  &  &  \\
 \hline
\end{tabular}
\vspace{-0.3 cm}
\label{tab.contrast}
\end{table} 

{For WET systems, the transmit signals carry energy. Hence, the energy harvester (EH) at the receiver harvests RF energy from the received signals. The EH relies on an antenna and a rectifier, which converts the RF signal power into the direct current (DC) by relying on a non-linear mapping characteristic \cite{hu2020energy}.}
Obviously, the specific characteristics of EHs have substantial impact on the WET performance at the receiver. Therefore, it is crucial to accurately model the nonlinear nature of the energy harvesting process precisely. Varasteh \textit{et al.} \cite{varasteh2020learning} proposed a pair of analytical EH models for low and high RF input power, respectively. However, some practical hardware impairments, such as the impedance mismatch and the non-ideal nature of the low-pass filters, were hard to model accurately. Accordingly, they proposed to characterise the EH model by a DNN and they investigated the IDEN performance in an end-to-end manner for the first time.

As one of the most important functions of an end-to-end communication system, channel decoding has a beneficial impact on the BER performance \cite{8936409}. Polar codes have been adopted in the 5G New Radio (NR) control channel as a benefit of its good performance at short block-length. Hence, many researches on polar code were conducted for further improvement, e.g., polar code design adapted to multiple fast fading channels \cite{9130148niu} and soft list polar decoding for multiple-input multiple-output (MIMO) system \cite{9185047Xiang}. As a further advance, deep learning aided polar decoding design were conceived in \cite{8109997,teng2019low,9109744zhu}.
Specifically, Zhu \textit{et al.} \cite{9109744zhu} designed a residual neural network decoder for polar codes, where  a denoising module based on residual learning was appended before the neural network.
{
In 3GPP Release 15 [15], the cyclic redundancy check-assisted successive cancellation list (CA-SCL) algorithm is standardized as the polar decoder because of its superiority in error correction. However, the BP algorithm achieves lower latency than the SCL due to its parallel structure. Due to the drawback of slow convergence and inferior error correction of the BP algorithm, the DNN based BP decoders are proposed to overcome these problems.} For instance,
Xu \textit{et al.} \cite{8109997} proposed a novel DNN based polar decoder, which reduced the latency and complexity compared to the conventional belief propagation (BP) based method, while a recurrent neural network (RNN)-aided polar decoder was proposed in \cite{teng2019low}, which required reduced memory without substantial performance erosion.


However, in existing systems, typically a single functional module (e.g   modulation/demodulation, EH or channel decoder) is implemented by DNN in isolation, which merely optimizes a single module, but fails to achieve globally optimal performance. Moreover, the application of polar codes in 5G demonstrates its practical significance, while the benefits of polar codes in the existing literature of IDEN have been overlooked, even though they are capable of substantially improving the WIT performance. Hence, harnessing them in IDEN systems is also expected to improve the WET performance, since we may be able to allocate more communication resources to WET services. Therefore, it is essential to consider end-to-end design of a polar coded IDEN system. 

Against this background, our main contributions are totally and explicitly contrasted to the existing literature in Table \ref{tab.contrast} at a glance and they are summarized in more detail as follows:
\begin{itemize}
  \item We conceive an end-to-end polar-coded IDEN system, where the polar code is harnessed both for data and energy transmission. The original functional modules of modulation, demodulation, EH and polar decoding are replaced by DNNs, which are jointly optimized to achieve an improved IDEN performance.
  \item By exploiting the similarities between the polar code's graph-based representation and the neural network connections, we formulate a DNN-aided BP based polar decoder. In contrast to \cite{8109997}, this decoder is designed for satisfying both the WIT and WET requirements, minimizing the BER while satisfying the energy harvesting requirement and the transmit power budget of the IDEN system designed. 
\end{itemize}
{Our proposed system provides a gain of almost $14$ dB in comparison to the traditional system at the target BER of $10^{-3} $ at $22$ dB with the aid of $3$ BP iterations and $\lambda=0.01$ for transmission over a Rayleigh fading channel.
}

\begin{figure}[t]
  \centering
  \includegraphics[width=3in]{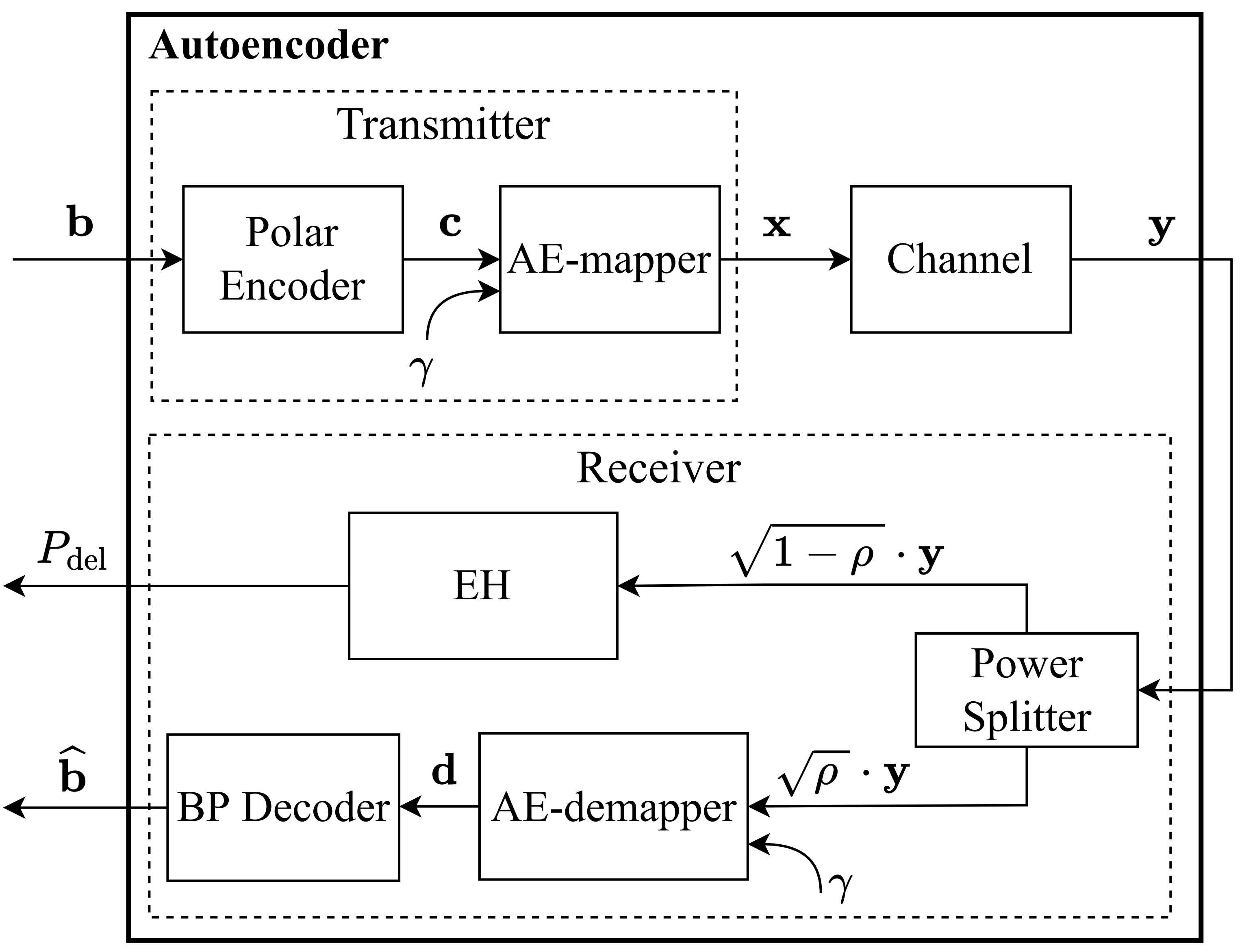}\\
  \caption{An end-to-end polar-coded IDEN system.}\label{fig.systemmodel}
  \vspace{-0.4 cm}
  \end{figure}

The rest of this paper is organised as follows. Our system model is described in Section II, while our optimization problem and the corresponding solution is detailed in Section III. After providing our simulation results in Section IV, we finally conclude in Section V.
  
\section{System Model}
\subsection{Transmitter}
The transmitter is constituted by a polar encoder and an AE mapper, as illustrated in Fig. \ref{fig.systemmodel}. 
\subsubsection{Polar Encoder}
A $K$-bit information sequence $\mathbf{b}$ is first polar-coded into an $N$-bit coded bit sequence.  To obtain an $(N,K)$ polar code, we  assign the information bits in $\mathbf{b}$ to the $K$ most reliable ``sub-channels`` out of the total of $N$ ``sub-channels``. The remaining $(N-K)$ bits are referred to as zero-valued frozen bits and they are assigned to the $(N-K)$ less reliable sub-channels. {The sub-channel reliability sequence we implemented is proposed in \cite{3rd2021technical}.}  Note that the positions and values of frozen bits are known by both the polar encoder and the decoder. Polar encoding is performed based on the combined information \& frozen bit sequence $\mathbf{u}$ having $N$ bits in total. The output $\mathbf{c}$ of the polar encoder is obtained as 
\begin{equation}
  \mathbf{c}=\mathbf{u}\textbf{G}_{N}=\mathbf{u}\mathbf{F}^{\otimes{n}}\mathbf{B}_{N},
\end{equation}
where $\textbf{G}_N$ is the generator matrix, while $\textbf{B}_{N}$ is the bit-reversal permutation matrix{, which is harnessed for simplifying the design of the decoder \cite{arikan2009channel}}. Furthermore, the symbol $\otimes$ denotes the Kronecker product and $\textbf{F}^{\otimes{n}}$ is the $n$-th Kronecker power of { $\textbf{F}= \begin{bmatrix}
\begin{smallmatrix}
 1 & 0 \\
 1 & 1
\end{smallmatrix}
\end{bmatrix}$ } associated with $n = \log_2 N$.

\subsubsection{AE-Mapper}
\begin{figure}
	\centering
	\subfigure[]{
	\label{fig.transmitter}
		\includegraphics[width=1.7in]{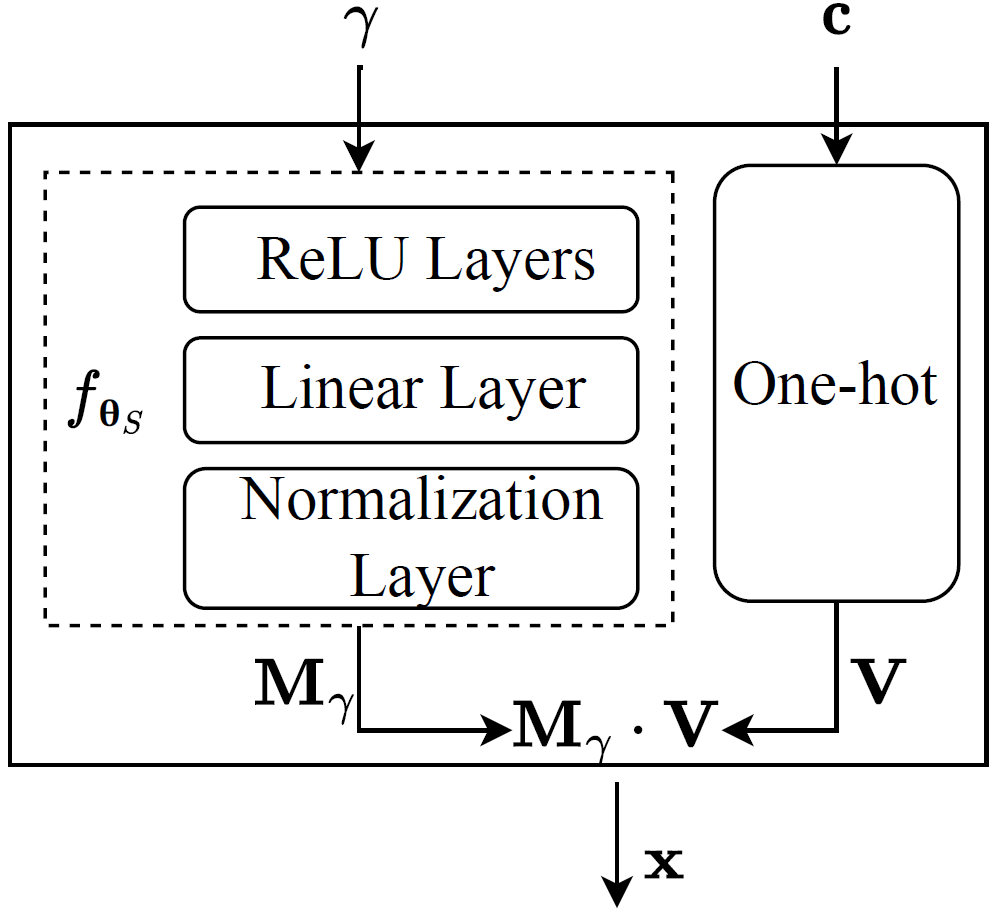}}
	\subfigure[]{
	\label{fig.receiver}
	\includegraphics[width=1.7in]{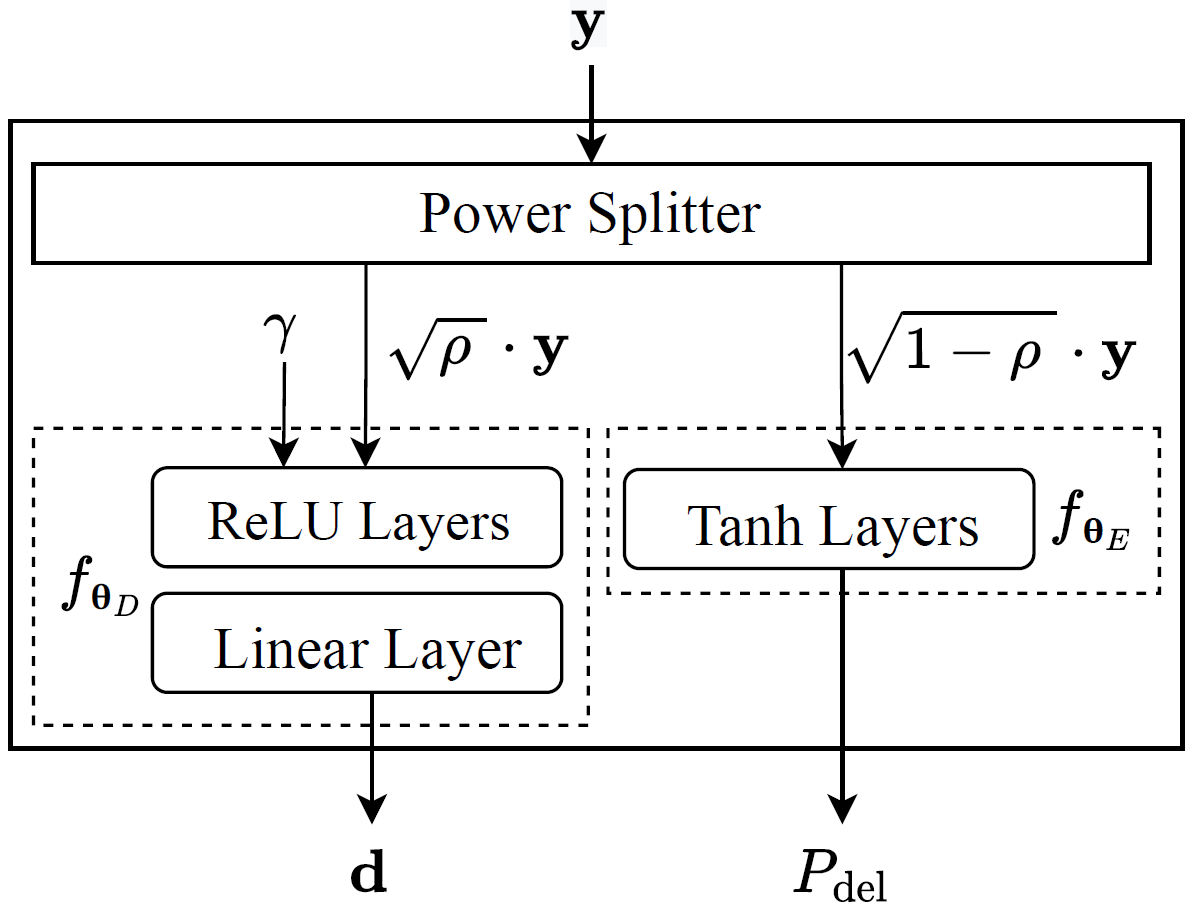}}
	\caption{A generic architecture of (a) AE-Mapper and (b) AE-Demapper and EH.}
	\label{fig.structure}
	\vspace{-0.4 cm}
\end{figure}

The AE-mapper performs the modulation function with the binary bit sequence as its input and modulated symbol as its output. A generic architecture of the AE-mapper is portrayed in Fig.~\ref{fig.transmitter}, where the channel signal-to-noise (SNR) $\gamma$ together with the polar encoded bit vector $\mathbf{c}$ are input to the AE-mapper. The AE-mapper includes a fully-connected $I^{\textrm{S}}$-layer DNN $f_{\pmb{\theta}_{\textrm{S}}}$, with the last layer being the so-called normalization layer for satisfying the transmit power constraint, as shown in Fig. \ref{fig.transmitter}.
One-hot mapping is applied to $\mathbf{c}$. Given the modulation order $M$, we have a matrix of dimension $\mathbf{V}\in \mathbb{C}^{M \times N/\log_2 M}$, where the $n$-th column represents a one-hot vector $\mathbf{v}_n \in \mathbb{C}^{M \times 1}$. Given a channel SNR $\gamma$, the DNN $f_{\pmb{\theta}_{\textrm{S}}}$ learns to constitute an $M$-ary constellation $\mathbf{M}_{\gamma} \in \mathbb{C}^{M \times 2}$, whose two columns represent the real and imaginary parts of the $M$ constellation points, respectively.
The function $f_{\pmb{\theta}_{\textrm{S}}}$ of the DNN can be expressed as
\begin{align}
\mathbf{M}_{\gamma}&=f_{\pmb{\theta}_{\textrm{S}}} (\gamma) \nonumber \\
& =f_{\textrm{norm}} \left(\mathbf{W}^{(\textrm{S})}_{I_{\textrm{S}}}\left [  \cdots f_{\textrm{ReLU}}(\mathbf{W}^{(\textrm{S})}_1 \gamma+\mathbf{b}^{(\textrm{S})}_1)\cdots\right ]+\mathbf{b}^{(\textrm{S})}_{I_{\textrm{S}}} \right), 
\end{align}
where $f_{\textrm{norm}}(\cdot)$ and $f_{\textrm{ReLU}}(\cdot)$ represent the normalization and the ReLU activation functions, respectively, while the trainable weight and bias parameters $\mathbf{W}^{(\textrm{S})}_i$ and $\textbf{b}^{(\textrm{S})}_i$ for $\forall i= 1, \dots, I_{\textrm{S}}$ are collected in the set $\pmb{\theta}_{\textrm{S}}$.

To obtain the modulated symbols $\mathbf{x}$, the constellation set $\mathbf{M}_{\gamma}$ is multiplied by the matrix $\mathbf{V}$ to map the one-hot vectors to the constellation points in $\mathbf{M}_{\gamma}$. Upon considering the $n$-th symbol as an example, the real and imaginary parts of the complex baseband symbol $x_n$ are obtained by multiplying the constellation matrix $\mathbf{M}_{\gamma}$ with the one-hot vector $\mathbf{v}_n$, which can be expressed as 
\begin{equation}\label{equ.map}
  [\Re(x_n)~\Im(x_n)]=\textbf{v}_n^T\cdot \mathbf{M}_{\gamma}.
\end{equation}
The resultant modulated symbol vector $\mathbf{x}=[x_1, \dots, x_{N/\log_2 M}]^T$ is then transmitted.

\subsection{Receiver}
As illustrated in Fig. \ref{fig.systemmodel}, the receiver consists of a power splitter, an AE-demapper, an EH and a BP decoder. The signal $\mathbf{y} \in \mathbb{C}^{N/\log_2 M \times 1} $ impinging at the single-antenna receiver can be expressed as
\begin{equation}
  \mathbf{y}=\mathbf{h}\odot\mathbf{x}+\mathbf{n},  
\end{equation}
where $\mathbf{h} \in \mathbb{C}^{N/\log_2 M \times 1}$ represents the channel coefficients, and $\odot$ denotes the element-wise multiplication. We assume encountering an uncorrelated Rayleigh fading channel, where $\mathbf{h}$ follows a complex Gaussian distribution $\textbf{h}\sim \mathcal{CN}(0,1)$, while the additive white Gaussian noise (AWGN) $\textbf{n}\in \mathbb{C}^{N/\log_2 M \times 1}$ follows $\textbf{n}\sim \mathcal{CN}(0,\sigma^2)$. Since the normalization layer of the AE-mapper ensures that $\mathbb{E}\left[ \|\pmb{x}\|^2 \right]=1$ and $2\sigma^2$ is the complex noise variance,  $\sigma^2$ can be expressed as $\sigma^2=\frac{\mathbb{E}\left[ \|\pmb{x}\|^2 \right]}{2\gamma}=\frac{1}{2\gamma}$.

\subsubsection{Power Splitter}
The received signal $\mathbf{y}$ is firstly input to the power splitter, which can divide the input signal into two branches according to the specified energy ratio. As shown in Fig. \ref{fig.systemmodel}, the parameter $0\leq \rho\leq 1$ denotes the power splitting factor that determines the energy ratio.
Then these two branches are forwarded to the AE-demapper and EH, respectively.

\subsubsection{AE-Demapper}\label{sec.receiver}
A portion of the received signal given by $\sqrt{\rho}\mathbf{y}$ is then fed into AE-demapper for demodulation, where the output $\mathbf{d}$ represents the prediction of the encoded sequence $\mathbf{c}$. The AE-demapper employs a $I_{\textrm{D}}$-layer DNN $f_{\pmb{\theta}_{\textrm{D}}}$ relying on the ReLU and on the linear activation functions for recovering the received symbol, as shown in Fig. \ref{fig.receiver}. The action of this DNN $f_{\pmb{\theta}_{\textrm{D}}}$ can be formulated as
\begin{align}
\mathbf{d}= & f_{\pmb{\theta}_{\textrm{D}}}(\mathbf{y})\nonumber \\
=& \mathbf{W}^{(\textrm{D})}_{I_{\textrm{D}}}\left [\cdots f_{\textrm{ReLU}}\left (\mathbf{W}^{ (\textrm{D})}_1  \mathbf{y}+\mathbf{b}^{(\textrm{D})}_1  \right ) \cdots \right ] +\mathbf{b}^{(\textrm{D})}_{I_{\textrm{D}}}  ,
\end{align}
where $\mathbf{W}^{(\textrm{D})}_i$ and $\textbf{b}^{(\textrm{D})}_i$ are collected into the parameter set $\pmb{\theta}_{\textrm{D}}$ denoting the weight and bias of the $i$-th layer in the DNN $f_{\pmb{\theta}_{\textrm{D}}}$ for $\forall i= 1, \dots, I_{\textrm{D}}$.

\subsubsection{EH}\label{sec.EH}
The remaining portion of the received signal, namely $\sqrt{1-\rho}\mathbf{y}$ flows into the EH. The harvested direct-current (DC) power is $P_{\textrm{del}}$, while the corresponding input RF power is $P_{\textrm{in}}$. The relationship between the input RF power and the output DC power can be modeled by a $I_{\textrm{E}}$-layer DNN $f_{\pmb{\theta}_{\textrm{E}}}$, where $\pmb{\theta}_{\textrm{E}}$ is the parameter set, as proposed in \cite{varasteh2020learning}. The function $f_{\pmb{\theta}_{\textrm{E}}}$ is formulated as
\begin{align}\label{equ.EH}
  P_{\textrm{del}}=&f_{\pmb{\theta}_\textrm{E}}(P_{\textrm{in}}) \nonumber \\
  =&f_{\textrm{tanh}}(\mathbf{W}^{(\textrm{E})}_{I_{(\textrm{E})}}\left [\cdots f_{\textrm{tanh}}(\mathbf{W}^{(\textrm{E})}_1\cdot P_{\textrm{in}}+\mathbf{b}^{(\textrm{E})}_1)\cdots \right ]+\mathbf{b}^{(\textrm{E})}_{I_{(\textrm{E})}}),
\end{align}
where we have $P_{\textrm{in}}=(1-\rho)\|\pmb{y}\|^2$ and $f_{\textrm{tanh}}(\cdot)$ represents the tanh function, while  $\textbf{W}^{(\textrm{E})}_i$ and $\textbf{b}^{(\textrm{E})}_i$ represent the weight and bias of the $i$-th layer in the DNN $f_{\pmb{\theta}_{\textrm{E}}}$ for $\forall i= 1, \dots, I_{\textrm{E}}$, respectively. Note that the EH model is trained separately in advance, using a nonlinear regression algorithm. Then the well-trained model operates as a fixed module in our system during the global training, without any further adjustment.

\begin{figure}
  \centering
  \includegraphics[width=3.1in]{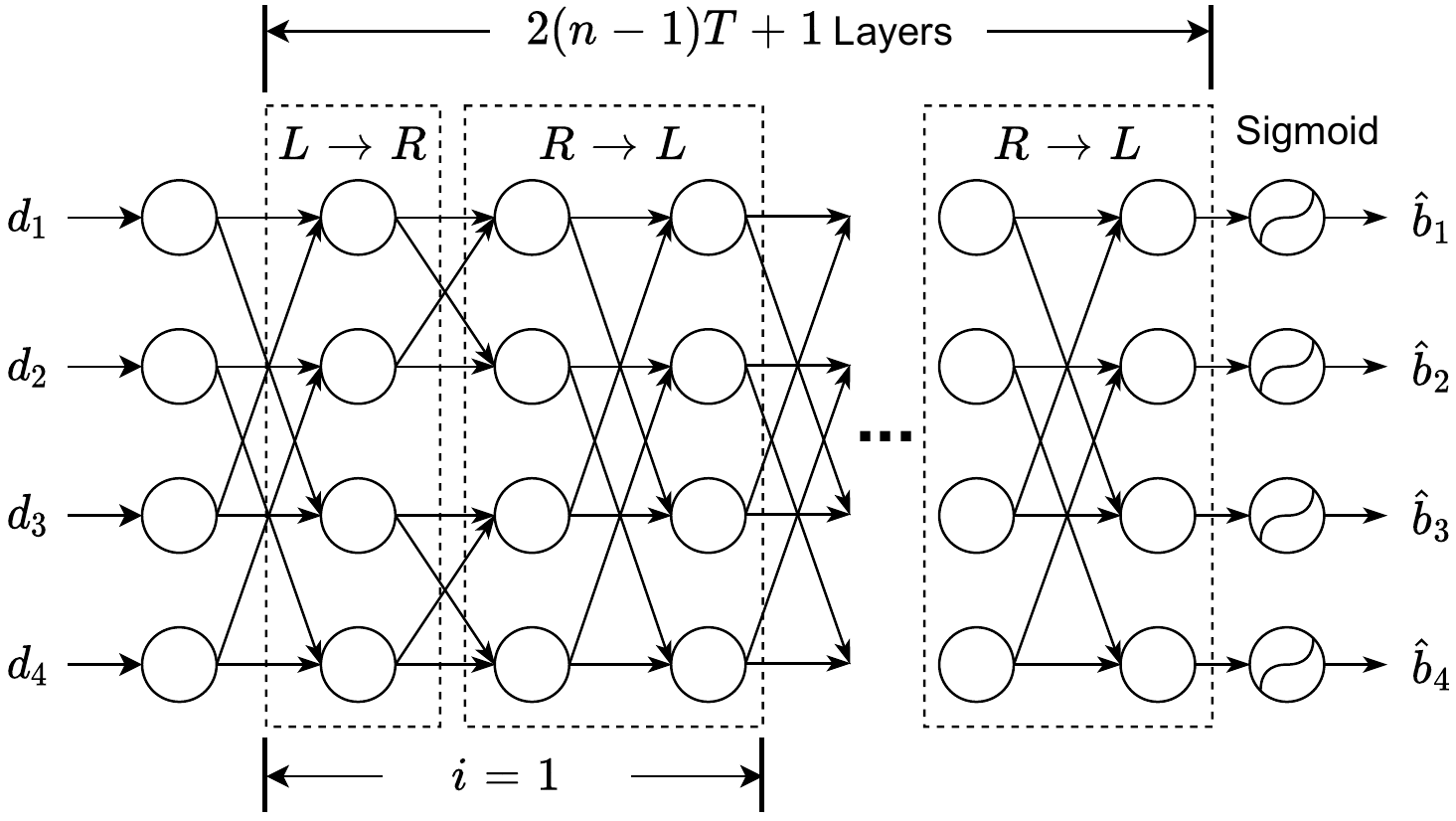}\\
  \caption{An example of DNN-based BP decoding with $N=4$.}\label{fig.BP}
  \vspace{-0.2 cm}
\end{figure}


\subsubsection{BP Decoder}\label{sec.decoder}
\begin{figure}[t]
  \centering
  \includegraphics[width=3in]{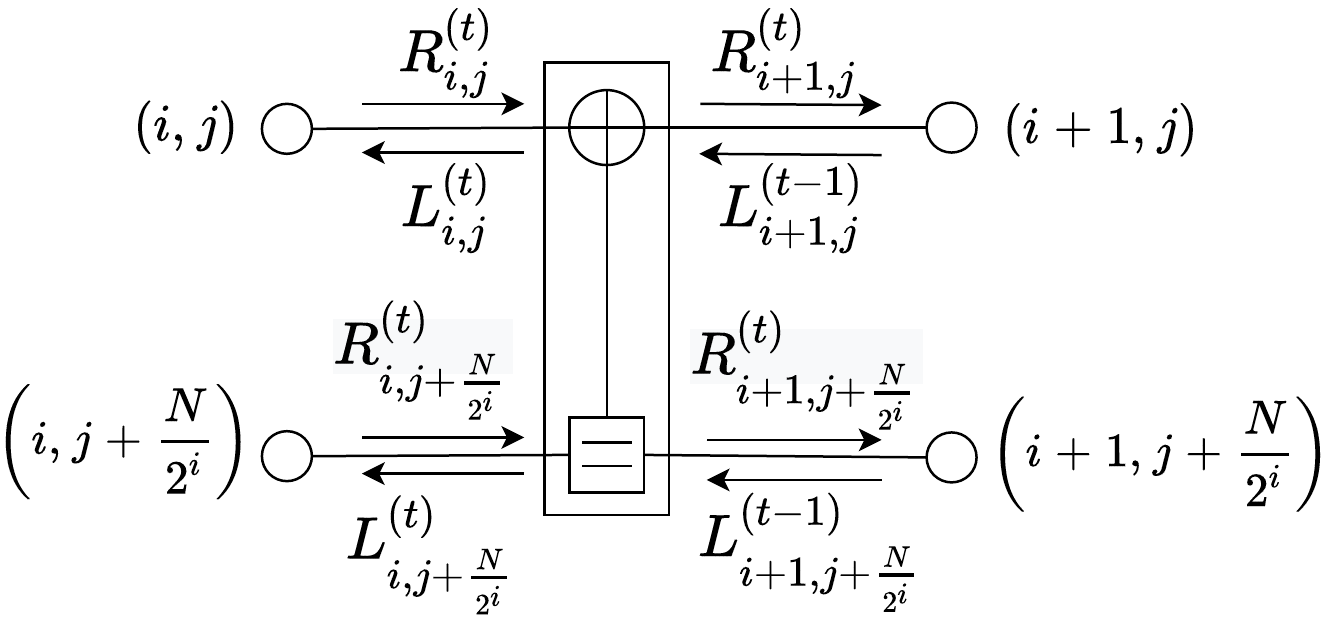}\\
  \caption{Processing element of the polar code graph.}\label{fig.PE}
  \vspace{-0.2 cm}
\end{figure}

After obtaining the demodulated vector $\mathbf{d}$, the DNN-aided BP algorithm processes the logarithmic likelihood ratios (LLRs) for carrying out channel decoding and outputs the prediction $\hat{\textbf{b}}$ of the original bits $\textbf{b}$. 
The conventional scaled BP decoder is replaced in our system by a multi-layer partially-connected DNN \cite{8109997}, where the connections between two layers correspond to those in the polar code's factor graph, as exemplified in Fig. \ref{fig.BP}. Generally, for a polar code of length $N$, its polar factor graph has $\log_2 N$ stages, which corresponds to a $[\log_2 N+1]$-layer DNN associated with $N(\log_2 N+1)$ neurons in total. Each layer has $N$ neurons. As the decoding iteration index $t=1,\cdots,T$ increases, the DNN expands by repeating the initial $\log_2 N$ number of stages. Specifically, the BP decoding process having $T$ iterations is represented by $[2(\log_2 N-1)T+1]$ hidden layers in the process of completing the left-to-right  $(L\rightarrow R)$ and right-to-left $(R\rightarrow L)$ LLR propagation, as illustrated in Fig. \ref{fig.BP}. 
The updates of the left-to-right LLR $R_{i,j}^{(t)}$ and the right-to-left LLR $L_{i,j}^{(t)}$ at the $t$-th iteration are formulated as
\begin{equation}\label{equ.scaledbp}
  \left\{
             \begin{array}{lr}
             L_{i,j}^{(t)}=\alpha_{i,j}^{(t)}\cdot g(L_{i+1,j}^{(t-1)},L_{i+1,j+N/{2^i}}^{(t-1)}+R_{i,j+N/{2^i}}^{(t)}),  \\
             L_{i,j+N/{2^i}}^{(t)}=\alpha_{i,j+N/{2^i}}^{(t)}\cdot g(R_{i,j}^{(t)},L_{i+1,j}^{(t-1)})+L_{i+1,j+N/{2^i}}^{(t-1)}, \\
             R_{i+1,j}^{(t)}=\beta_{i+1,j}^{(t)}\cdot g(R_{i,j}^{(t)},L_{i+1,j+N/{2^i}}^{(t-1)}+R_{i,j+N/{2^i}}^{(t)}), \\
             R_{i+1,j+N/{2^i}}^{(t)}=\beta_{i+1,j+N/{2^i}}^{(t)}\cdot g(R_{i,j}^{(t)},L_{i+1,j}^{(t-1)})+R_{i,j+N/{2^i}}^{(t)},
             \end{array}
  \right.
  \end{equation}
where we have $g(a,b)\approx \textrm{sign}(a)\textrm{sign}(b)\min(|a|,|b|)$, and $\alpha_{i,j}^{(t)}$ as well as $\beta_{i,j}^{(t)}$ are the right-to-left and the left-to-right scaling parameters of the $j$-th neuron at $i$-th stage during the $t$-th iteration, respectively. In the DNN-aided decoder, the basic computation unit termed as a ``processing element'' is composed of connected neurons as shown in Fig. \ref{fig.PE}. The LLRs update throughout this process according to Eq. \eqref{equ.scaledbp}.

To ensure that the output falls into the range of $[0,1]$, the classic sigmoid activation function $f_{\textrm{sigmoid}}$ is employed by the last layer of the DNN. 
The function of the DNN aided BP decoder $f_{\pmb{\theta}_{\textrm{BP}}}$ can be expressed as
\begin{align}
    \hat{\textbf{b}}=f_{\pmb{\theta}_{\textrm{BP}}}(\textbf{d}).
\end{align}
Note that the structure of the DNN-aided BP decoder $f_{\pmb{\theta}_{\textrm{BP}}}$ follows the design guidelines of \cite{8109997}.

{Given the SNR-dependent characteristics of the system, the neural network parameters are susceptible to the channel conditions, which has a substantial impact on the communication performance. Therefore, in order to enhance the adaptability of our system to time-variant communication scenarios and reduce the offline training time, we train it for multiple SNRs within a complete training process. We select three SNRs with an appropriate spacing of $2$ dB as a training SNR set to avoid the SNR range becoming too wide, which may result in poor performance at some specific SNR levels.
}

\section{DNN based End-to-End Design}\label{sec.train}
In this section, the end-to-end optimization problem is formulated for our IDEN system, followed by our end-to-end training example for characterizing the overall process.

\subsection{Optimization Problem}\label{subsec.op}
We aim for satisfying the energy harvesting requirement $P_{\textrm{del}}$, while minimizing the BER performance. Hence, the  optimization problem of our AE architecture can be formulated as
\begin{align}\label{pf-1}
    \textrm{(P1):}\underset{\pmb{\theta}_\textrm{S},\pmb{\theta}_\textrm{D},\pmb{\theta}_{\textrm{BP}},\rho }{\text{min}}&  \textrm{E}  \Big[ \underbrace{\sum_{n=1}^N{\left( b_n\log \hat{b}_n+\left( 1-b_n \right) \log \left( 1-\hat{b}_n \right) \right)}}_{\textrm{WIT Part}} \nonumber \\
    &+\underbrace{\frac{\lambda}{P_{\textrm{del}}}}_{\textrm{WET Part}}   \Big ] 
    \\
    \tag{\ref{pf-1}{a}}\label{eq.Pdlconstraint}
    \text{s.t.}~ & (2), (5), (6) ~\textrm{and}~ (8) , \\
    \tag{\ref{pf-1}{b}}\label{eq.Ptrconstraint}
    &\|\pmb{x}\|^2\leq P_{\textrm{tr}},  
\end{align}
where $P_{\textrm{tr}}$ represents the transmit power constraint, and the bias parameter $\lambda$ characterizes the data vs. energy trade-off, which is introduced for striking a flexible trade-off between the communication and energy harvesting requirements. As seen in Eq. \eqref{pf-1}, the objective function (OF) is constituted by a pair of WIT and WET parts. The WIT part is characterized by the binary cross entropy (CE) between the original bits $b_n$ and its prediction $\hat{b}_n$, for $\forall n= 1, \dots, N$. The WET part in Eq. \eqref{pf-1} is jointly determined by both the bias parameter $\lambda$ and the harvested energy $P_{\textrm{del}}$. A larger $\lambda$ indicates that the IDEN has to harvest more energy at the cost of degraded BER performance and vice versa. Eq. \eqref{eq.Pdlconstraint} represents that the end-to-end IDEN process follows the mapping relationship of each neural network module, while Eq. \eqref{eq.Ptrconstraint} represents the transmit power constraint.

\begin{figure}[t]
	\centering  
	\subfigure[Conventional 8PSK]{
	\label{fig.original}
		\includegraphics[width=1.7in]{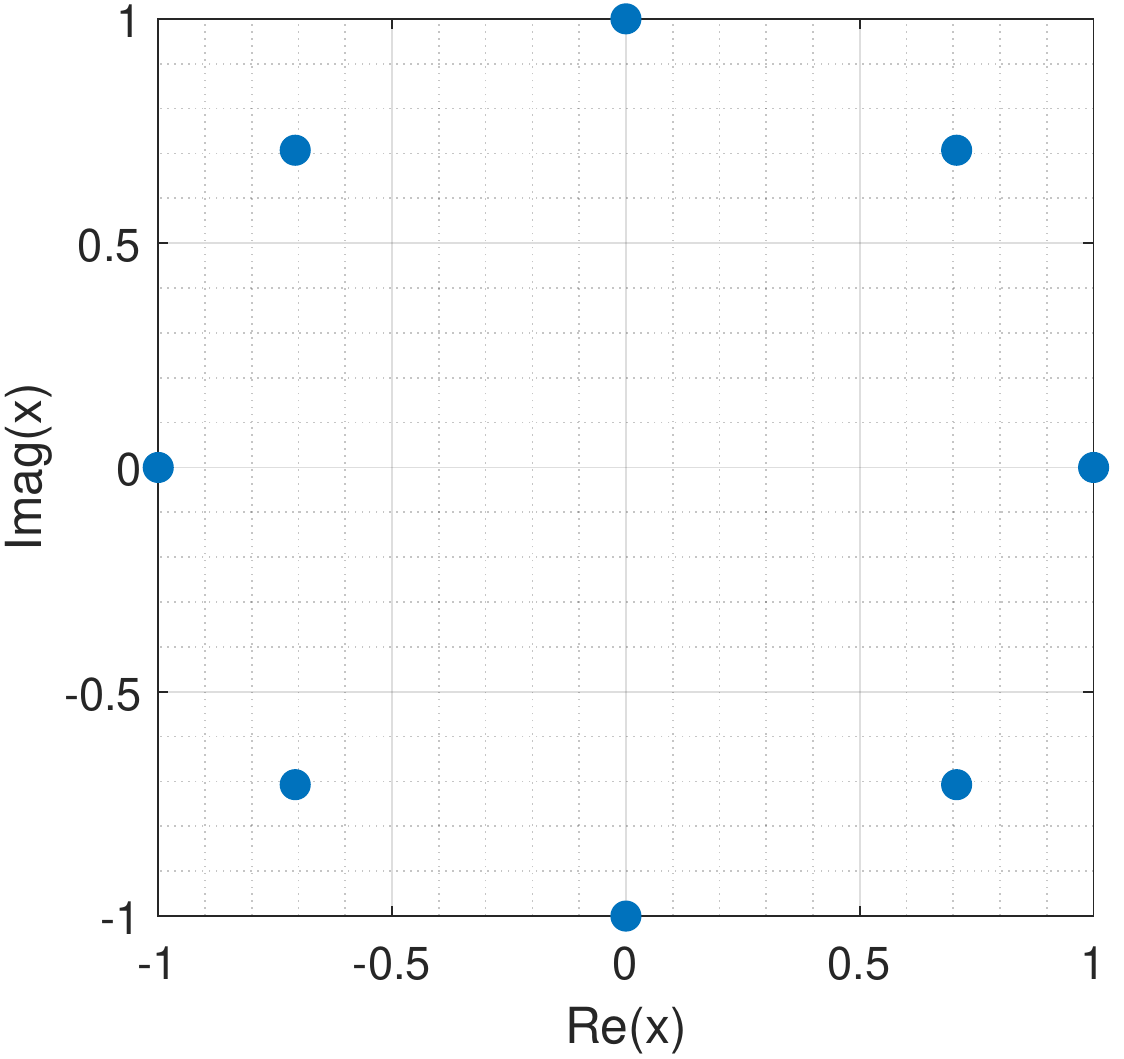}}
	\subfigure[AE-mapper, $\lambda=0$]{
	\label{fig.cons1}
		\includegraphics[width=1.7in]{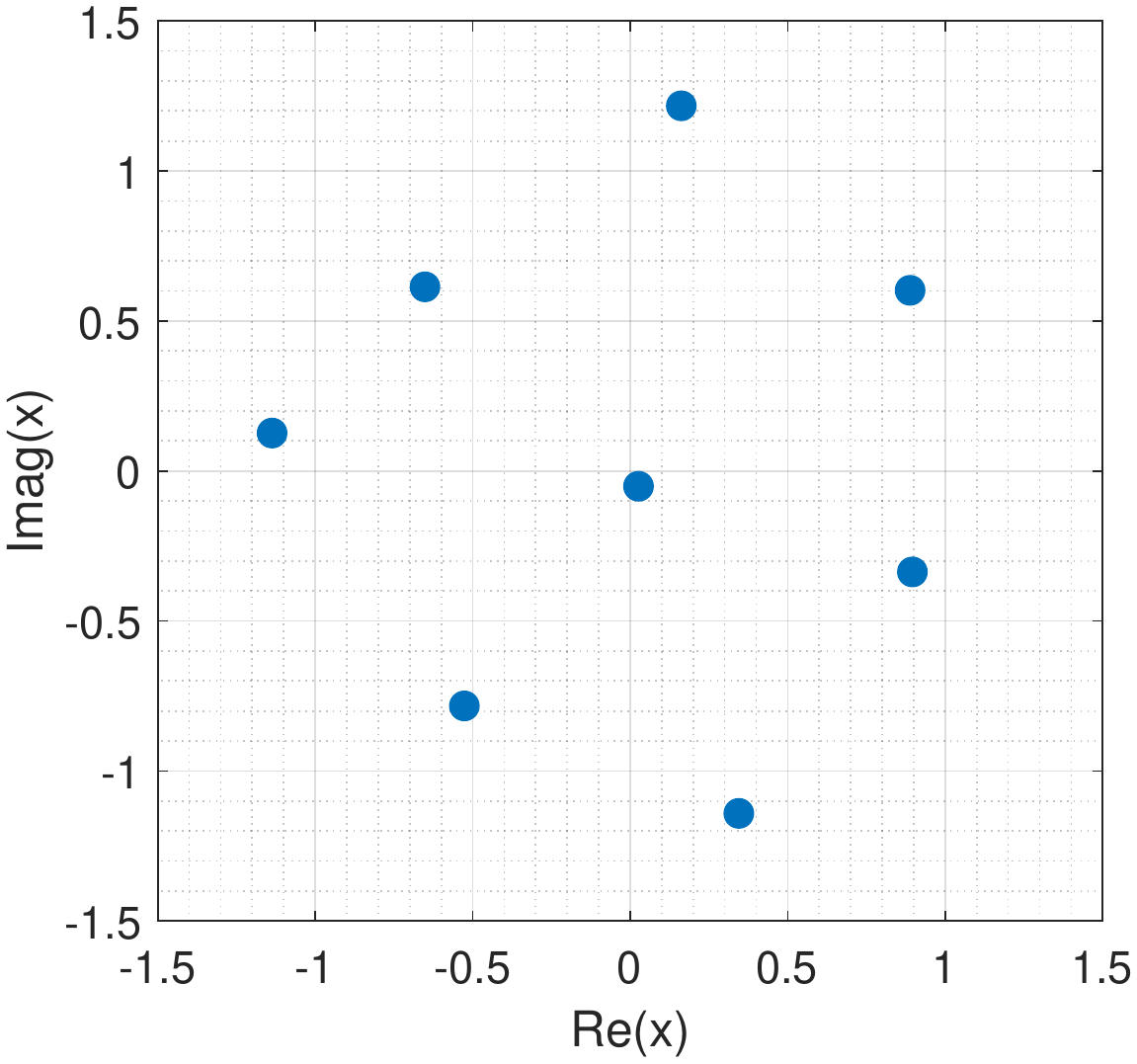}}
	\subfigure[AE-mapper, $\lambda=0.05$]{
	\label{fig.cons2}
	\includegraphics[width=1.7in]{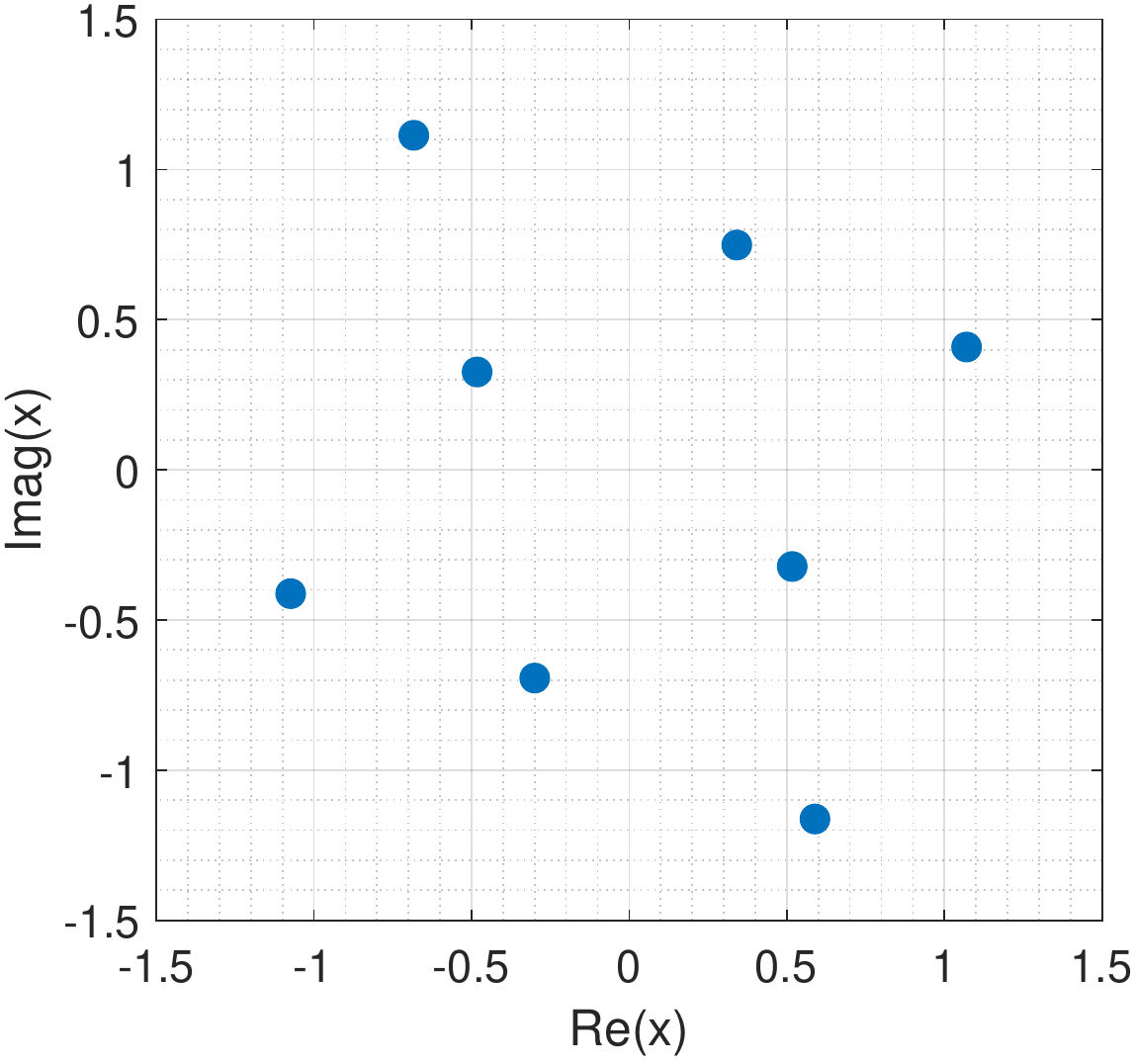}}
	\caption{Constellations of the conventional 8PSK and of the AE-mapper for different values of $\lambda$ at SNR = 8dB}
	\label{fig.constellation}
	\vspace{-0.3 cm}
\end{figure}

\subsection{End-to-End Training}
First, we randomly generate the information bit-vector $\textbf{b}$ and forward it to the polar encoder in order to output the coded bit vector $\textbf{c}$. The AE-mapper then maps the input coded bit vector $\textbf{c}$ onto a set of modulated complex-valued symbols  $\textbf{x}$. Then, these symbols $\textbf{x}$ are transmitted to the receiver through an AWGN/Rayleigh channel. After passing through the power splitter, the received signal are fed to the AE-demapper and EH of Fig. \ref{fig.receiver}. The EH collects the power of the signal portion of $\sqrt{1-\rho}\mathbf{y}$ and outputs the harvested energy $P_{\textrm{del}}$. After being processed by AE-demapper, the probability vector \textbf{d} is obtained. Finally, $\textbf{d}$ is forwarded to the BP decoder of Fig.\ref{fig.BP} and we get the predicted vector $\hat{\textbf{b}}$.
To evaluate the overall performance of this end-to-end communication system, we consider  both the WIT and WET, when designing the loss function. We adopt CE to express the difference of the final output $\hat{\textbf{b}}$ and the original input $\textbf{b}$, which represents the WIT loss, and introduce the parameter $\lambda$ together with the harvested energy $P_{\textrm{del}}$ to evaluate the WET loss. The loss function is formulated as follows:
\begin{align}\label{equ.loss}
  L(\textbf{b},\hat{\textbf{b}},\pmb{\theta})=&-\frac{1}{|\mathcal{B}|}\sum_{\textbf{b}\in |\mathcal{B}|}\left(\sum_{n=1}^N\left( b_n\log \hat{b}_n+( 1-b_n ) \right . \right .  \nonumber \\
  &\times 
  \left.\left.\log ( 1-\hat{b}_n ) \right)   +\frac{\lambda}{P_{\textrm{del}}(\textbf{b})} \right),
\end{align}
where $|\mathcal{B}|$ denotes the batch size of the training samples and $\pmb{\theta}$ represents a parameter set, which contains all the trainable parameters, including $\pmb{\theta}_\textrm{S},\pmb{\theta}_\textrm{D}$ and $\pmb{\theta}_{\textrm{BP}}$. In the IDEN system, we train and optimize the DNNs by minimizing the loss function, which corresponds to the OF of $\textrm{(P1)}$, as we summerized in section \ref{subsec.op}. The training process also obeys the constraints listed in the optimization problem.
All trainable parameters of the system are updated iteratively using the classic stochastic gradient descent (SGD) algorithm, which can be formulated as:
\begin{equation}\label{equ.sgd}
\setlength{\abovedisplayskip}{0cm}
  \setlength{\belowdisplayskip}{0cm}
  \pmb{\theta}_{T+1}=\pmb{\theta}_{T}-\delta\nabla L(\textbf{b},\hat{\textbf{b}},\pmb{\theta}_{T}),
\end{equation}
where $\delta$ denotes the learning rate and again, $t=1,\cdots,T$ denotes the iteration index of the parameter update process.

\begin{figure*}[t]
	\centering  
	\subfigure[]{
	\label{fig.R_iter}
		\includegraphics[width=2.9in]{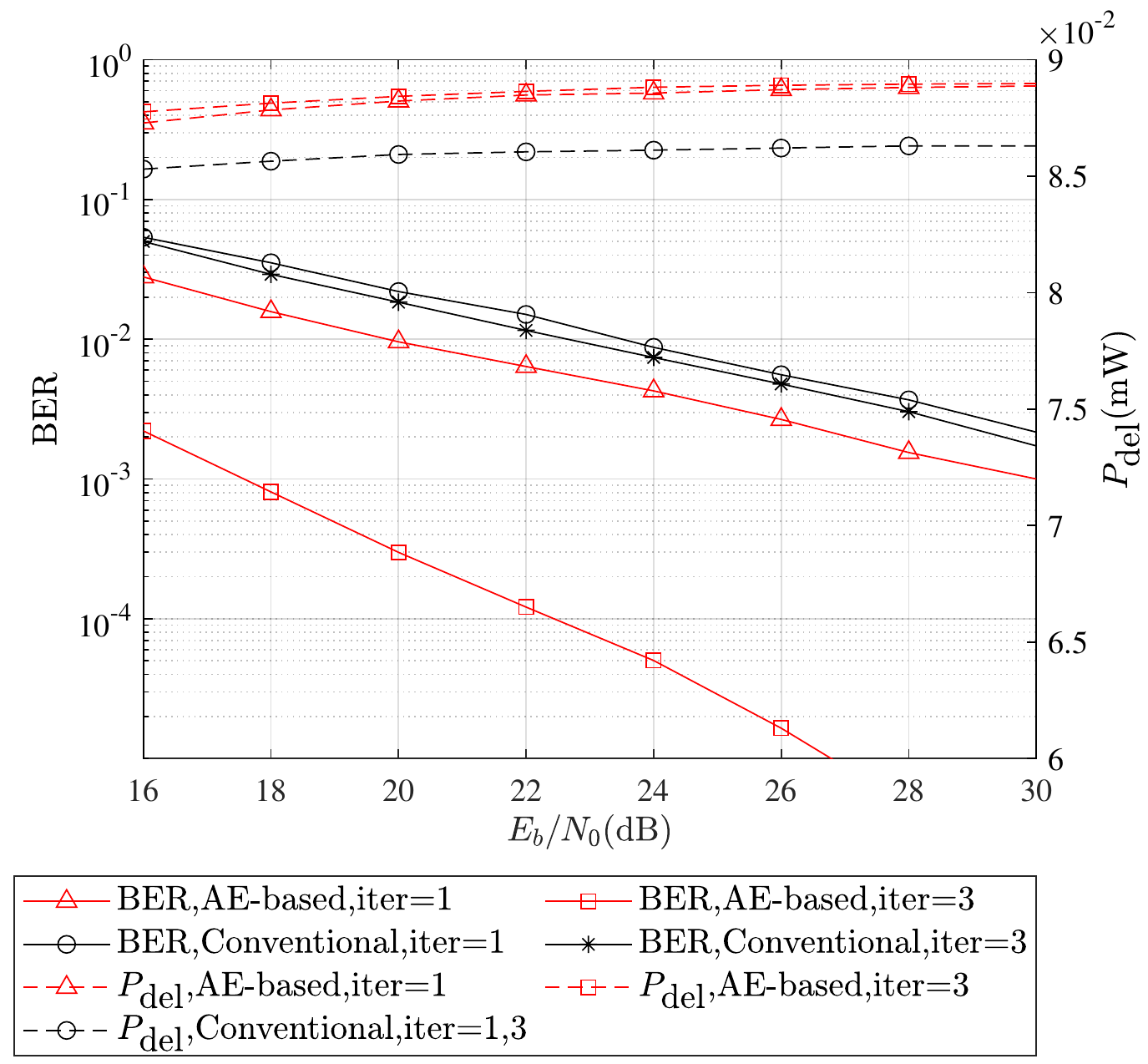}}
			\subfigure[]{
	\label{fig.R_lamda1}
	\includegraphics[width=2.9in]{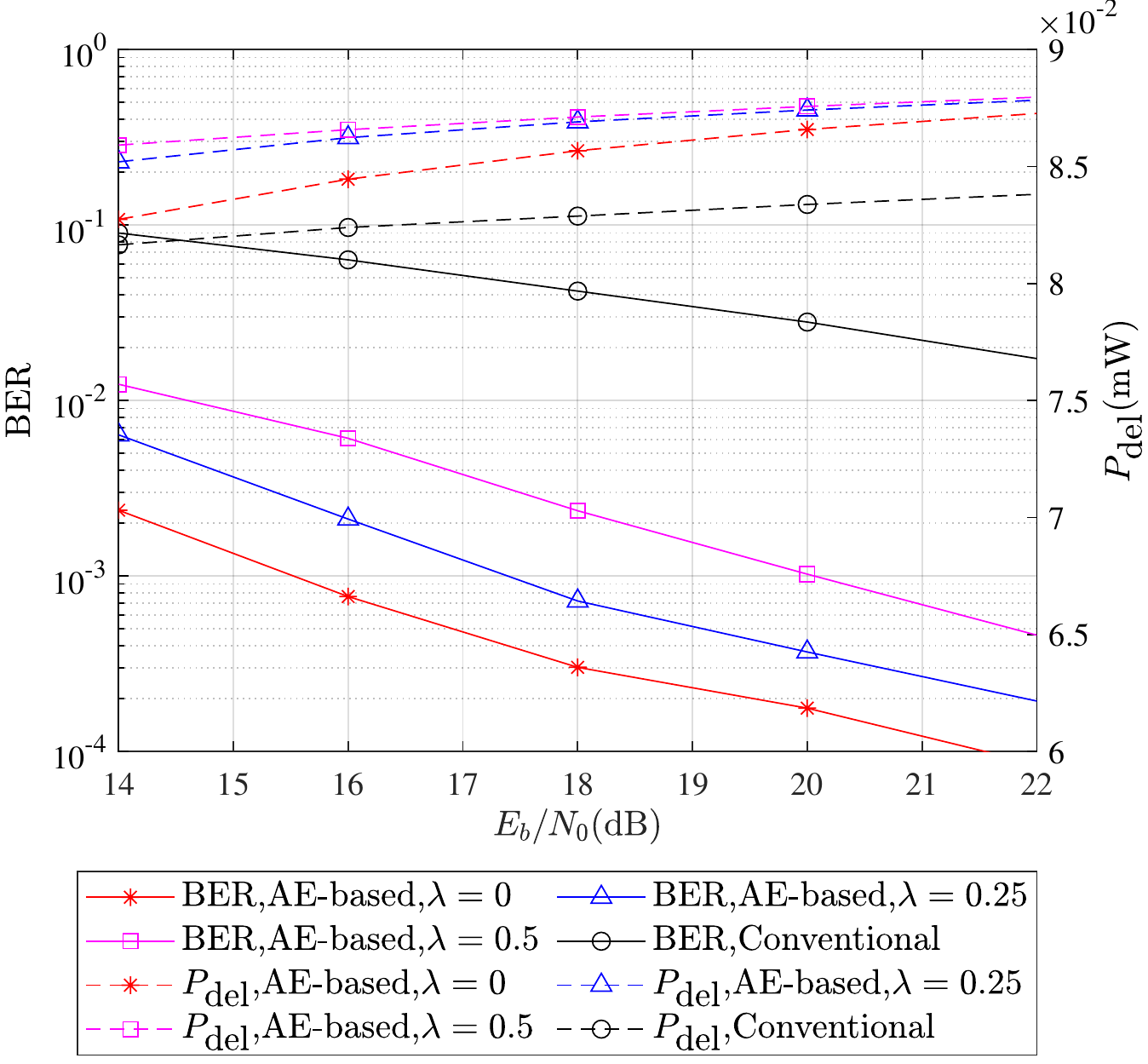}}
		\vspace{-0.2 cm}
	\caption{BER and energy harvesting performance of IDEN system over Rayleigh channel: (a) Performance of IDEN system and conventional system with $\lambda=0.01$, the noise power $P_{noise}=0.17$ dBm under different iterations; (b) Performance of the IDEN system with iteration=3 and different $\lambda$.}
	\label{fig.R_performance}
	\vspace{-0.3 cm}
\end{figure*}

{Compared to the isolated training model, the main drawback of the proposed joint training model is its complexity. In terms of the joint training model, the complexity of the end-to-end training is $\mathcal{O}((T+1)N\textrm{log}N+Q_m^2+2MQ_m+Q_d^2+Q_dlog_2M)$, which is more complex than the isolated training model associated with $\mathcal{O}((T+1)N\textrm{log}N)$. Here $Q_m$ and $Q_d$ represent the number of neurons in each layer of the AE-mapper and AE-demapper, respectively. }

\section{Simulation Results}

\begin{table}
\centering
{
\caption{Details Of The DNN of Each Module}
\begin{tabular}{|c|c|c|} 
 \hline
 Module & Layer & Size \\ \hline
 \multirow{3}*{AE-mapper}
  & ReLU & $128 \times 128$\\ 
 \cline{2-3}
  & Linear & $128 \times 16$\\ 
 \cline{2-3}
  & Normalization & \diagbox[height=7pt,innerrightsep=42pt,dir=SW]{}{}\\ 
 \hline
 \multirow{2}*{AE-demapper}
  & ReLU & $128 \times 128$ \\ 
 \cline{2-3}
  & Linear & $128 \times 3$\\ 
 \hline
 \multirow{1}*{BP Decoder} 
 & Non-fully connected layer & $64 \times 64$\\
 \hline
\end{tabular}
\vspace{-0.3 cm}
\label{tab.detail}
}
\end{table} 

In this section, we evaluate the performance of the proposed polar-coded IDEN system over both AWGN and Rayleigh channels. A polar code having $K=32$ and $N=64$ is employed.
We implement our system on TensorFlow 1.14. {The number of training epochs for each training SNR is set to $E=5$}, and in each epoch, the training samples are randomly generated with the mini-batch size being 1000. We use the Adam optimizer with a learning rate of $\delta = 0.005$. The training SNR ranges from 
16 dB to 30 dB for Rayleigh channels. The power splitting factor $\rho$ is set to $\frac{\sqrt{2}}{2}$ in our system, which means that the data and energy branches output by the power splitter have equal energy. {Moreover, the size and activation functions of layers in each DNN based module are summarized in Table \ref{tab.detail}. The AE-mapper and the AE-demapper rely on fully connected neural networks, where the activation functions for the hidden layers are either the ReLU function or the Linear function. By contrast, the BP decoder is formed by a non-fully connected neural network, whose number of layers depends on the index of BP iterations.}

We first investigate the constellation output by the AE-mapper. Fig. \ref{fig.constellation} shows substantial difference between the conventional $8$-PSK constellation and the output of our AE-mapper having $M=8$ at different values of the bias parameter $\lambda$ over AWGN channels. The constellation points in Fig. \ref{fig.original} are uniformly distributed on the circumference of a circle. The Euclidean distance between the adjacent constellation points is limited at a given power, which determines the BER performance. By contrast, the constellation points output by the AE-mapper have unequal distances, as shown in Fig. \ref{fig.cons1} and \subref{fig.cons2}, which depend on the parameter $\lambda$.
Moreover, the constellation associated with $\lambda=0.05$ in Fig. \ref{fig.cons2} is different from that of $\lambda = 0$ in Fig. \ref{fig.cons1}. This demonstrates that the data vs. energy trade-off of the IDEN receiver affects the optimal shape of the AE-mapper's output constellation. This is because we take $\lambda$ into consideration, while designing the loss function. As $\lambda$ increases, the WET part dominates the loss function, and hence the DNNs are mainly trained in order to improve the energy harvesting performance. The constellation of Fig. \ref{fig.cons1} is constructed for exclusively optimizing the WIT performance, while that of Fig. \ref{fig.cons2} strikes a compromise between the WIT and WET performance, which ensures that the amplitude of transmitted phasor is large enough. Therefore, the constellation Fig. \ref{fig.cons2} achieves a better WET performance than that of Fig. \ref{fig.cons1} .
		

Let us now compare the BER and the energy harvesting performance of the proposed DNN-aided IDEN system to that of the conventional $8$-PSK and BP decoding. These simulations are carried out using $M=8$ and $\lambda=0.01$ for a Rayleigh channel and either $1$ or $3$ BP iterations. Observe from Fig. \ref{fig.R_iter} that in the Rayleigh channel, the BER performance of both systems improves as we increase the number of iterations. However, the improvement attained by the DNN-aided system is more significant, than that of its conventional counterpart. This explicitly demonstrates our advantage of the DNN-aided scaled BP decoder, where the scaling parameters $\alpha_{i,j}^{(t)}$ and $\beta_{i,j}^{(t)}$ of the BP decoder achieve near-optimality after training for just a few epochs. 
The simulation results demonstrate that our IDEN system performs well in practical propagation environments in the face of both fading and noise.

To explore the impact of WET on WIT, the DNN-aided IDEN system having different data vs. energy demands $\lambda$ over Rayleigh channel is investigated in Fig. \ref{fig.R_lamda1}, where we have $M=8$ and $3$ BP iterations. Observe from Fig. \ref{fig.R_lamda1} that upon increasing $\lambda$ of the IDEN  receiver in the DNN-aided system, the energy curve is significantly shifted upward and the BER performance deteriorates, as expected. By contrast, since the $8$-PSK constellation is fixed and conventional design does not strike a tradeoff between the WIT and WET performance, the IDEN performance of the conventional system remains unchanged. Hence, it cannot achieve a satisfactory WET performance. The simulation results demonstrate the superiority of the DNN-aided system in terms of WET (e.g., the harvested energy increases by $2.5\times10^{-3}$ mW when $\lambda$ increases from 0 to 0.25 at SNR$=14$dB). Moreover, we can control the tradeoff between WIT and WET by adjusting the parameter $\lambda$ depending on the near-instantaneous demands. Furthermore, with the increase of $\lambda$, the amount of harvested energy decreases, which is in line with the non-linear relationship between the input RF power $P_{\textrm{in}}$ and the output DC power $P_{\textrm{del}}$.

\begin{figure}
  \centering
  \includegraphics[width=2.9in]{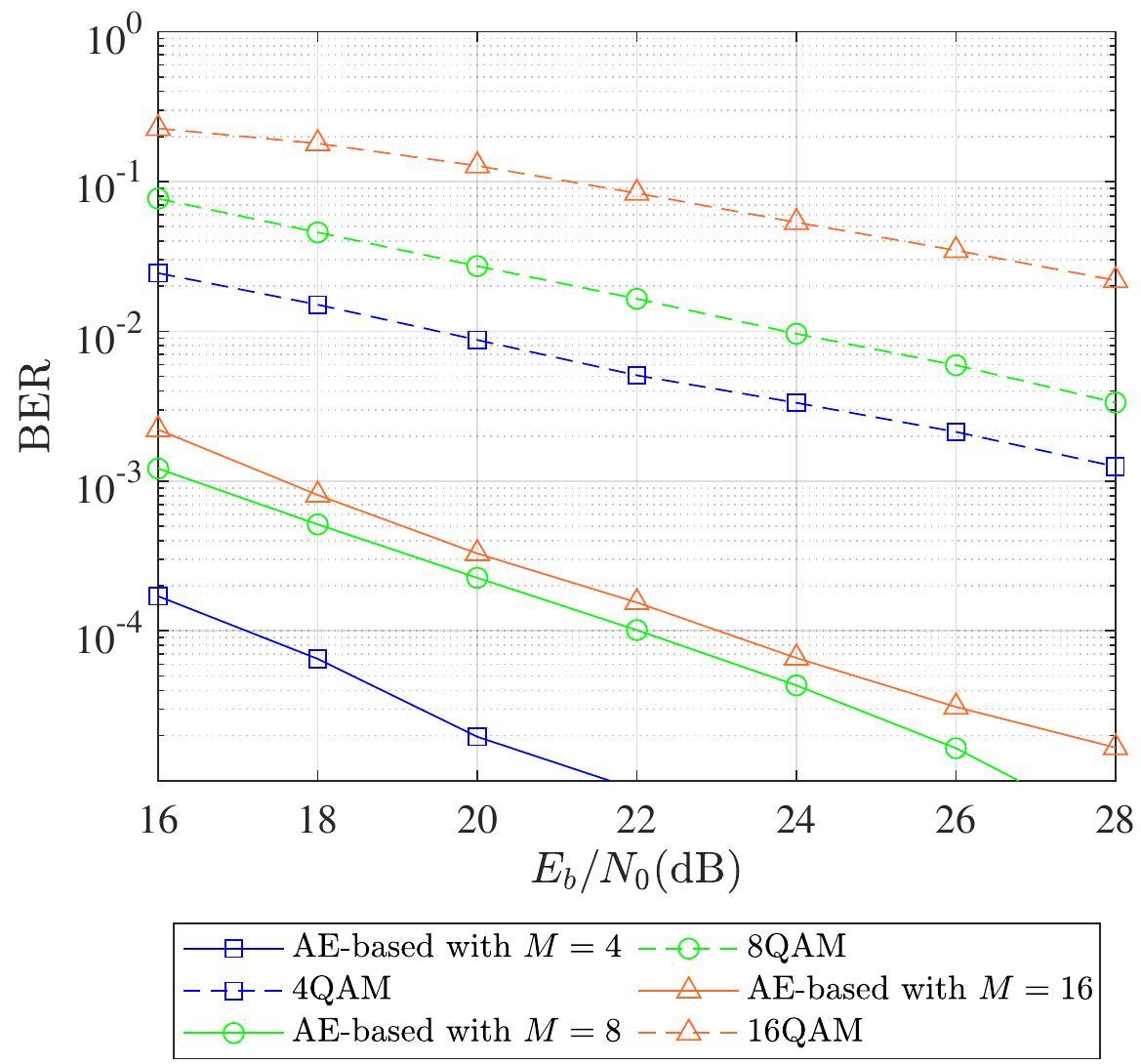}\\
  \caption{BER performance of IDEN system and traditional systems with $\lambda = 0.01$ and 3 BP iterations over Rayleigh channel.}\label{fig.revision}
  \vspace{-0.2 cm}
\end{figure}

{Given that 5G predominantly relies on QAM rather than PSK for modulation, we compared  the BER performance of our proposed DNN-aided IDEN system and the traditional modulation technologies, e.g., $4$-QAM, $8$-QAM and $16$-QAM, to provide a more comprehensive comparison. The simulations were carried out using $\lambda=0.01$ for a Rayleigh channel and 3 BP iterations. As shown in Fig.\ref{fig.revision}, our AE-aided modulation scheme outperforms its corresponding traditional counterparts at the same modulation order $M$. Due to the adaptability of AE, which can adjust its trainable parameters according to the time-variant channel conditions, the advantage of our IDEN system is clearly demonstrated.}

\section{Conclusions}
A DNN-aided polar-coded IDEN system was  proposed, which replaces the conventional functional modules by DNNs and characterises the whole system as an AE. All the DNNs can be trained in an end-to-end manner for the sake of jointly optimising the WET and the WIT performance. Our simulation results conducted in both AWGN and Rayleigh channels demonstrate the superiority of our data-driven end-to-end design over its conventional model-based counterpart in terms of both the BER and the energy harvesting performance.

\vspace{-0.1 cm}
\bibliographystyle{ieeetr}	
\bibliography{mm}

\end{document}